\documentstyle[12pt]{article}
\renewcommand{\theequation}{\arabic{equation}}
\renewcommand{\thesection}{\arabic{section}}
\renewcommand{\thefootnote}{\fnsymbol{footnote}}
\newcommand{\bea}{\begin{eqnarray}}
\newcommand{\ena}{\end{eqnarray}}
\newcommand{\vs}[1]{\vspace{#1 mm}}

\newcommand{\z}{\omega}

\newcommand{\PL}[1]{Phys.\ Lett.\ {\bf #1}}

\newcommand{\PR}[1]{Phys.\ Rev.\ {\bf #1}}
\newcommand{\PRL}[1]{Phys.\ Rev.\ Lett.\ {\bf #1}}

\newcommand{\PTPS}[1]{Prog.\ Theor.\ Phys.\ Suppl.\ {\bf #1}}
\newcommand{\AJ}[1]{Astrophys.\ J.\ {\bf #1}}

\newcommand{\EPJ}[1]{Eur.\ Phys.\ J.\ {\bf #1}}

\begin{document}
\noindent
\topmargin 0pt
\oddsidemargin 5mm

\begin{titlepage}
\setcounter{page}{0}
\begin{flushright}
September, 1999\\
OU-HET 326\\
hep-ph/9909415\\
\end{flushright}
\vs{10}
\begin{center}
{\Large{\bf Large CP Violation, Large Mixings of 
Neutrinos and  Democratic-type Neutrino 
Mass Matrix}}\\
\vs{10}
{\large  
Kiko Fukuura\footnote{e-mail address: 
 kiko@het.phys.sci.osaka-u.ac.jp}, 
Takahiro Miura\footnote{e-mail address: 
 miura@het.phys.sci.osaka-u.ac.jp}, 
Eiichi Takasugi\footnote{e-mail address: 
 takasugi@het.phys.sci.osaka-u.ac.jp} 
\\ and Masaki Yoshimura\footnote{e-mail address: 
 masaki@het.phys.sci.osaka-u.ac.jp}}
\\
\vs{8}
{\em Department of Physics,
Osaka University \\ Toyonaka, Osaka 560, Japan} \\
\end{center}
\vs{10}
\centerline{{\bf Abstract}}  
We propose a democratic-type neutrino mass matrix 
based on $Z_3$ symmetry. This mass matrix  
predicts the CP violation phase, $\delta=\pi/2$, and 
the mixing angle between the 
mass eigenstates $\nu_2$ and $\nu_3$, 
$\sin^2 \theta_{23}= \cos^2 \theta_{23}=1/2$  which is 
essential for the large atmospheric neutrino mixing  
between $\nu_\mu$ and $\nu_\tau$. In this model, the large 
CP violation effect may be expected. 
  
\end{titlepage}

\newpage
\renewcommand{\thefootnote}{\arabic{footnote}}
\setcounter{footnote}{0}

\section{Introduction}

The recent data on the atmospheric neutrino by 
Super-Kamiokande (Super-K)[1] show that the origin of the 
zenith angle dependence of neutrino flux 
is due to the oscillation between $\nu_\mu$ and 
$\nu_\tau$. The possibility of the $\nu_\mu$ 
oscillation to the sterile neutrino $\nu_s$ is 
almost excluded[1]. Also, the possibility of $\nu_\mu$ 
to $\nu_e$ oscillation is small[1] in accordance with 
the CHOOZ data[2]. The Super-K data is strengthened by 
the other data by MACRO[3] and Soudan 2[4] experiments. 
The preferable values of mass and mixing parameters are 
\bea
\sin^2 2\theta_{atm}=1.0\;,\; 
\Delta m^2_{atm}=3.5\times 10^{-3}{\rm eV^2}.
\ena
At 90$\%$ confidence level, the allowed region is 
$2\times 10^{-3}{\rm eV^2}
<\Delta m^2_{atm}<6\times 10^{-3}{\rm eV^2}$ and 
$\sin^2 2\theta_{atm}>0.85$. 

The situation of the solar neutrino problem is more involved. 
There are various solutions that  explain the 
absolute flux deficits by the Homestake[5], the Super-K[6], 
the GALLEX[7] 
and the SAGE[8] data, the small angle 
MSW solution ($\Delta m^2_{solar}={\rm a \; few}\times 
10^{-5}{\rm eV^2}$), the large angle MSW solution 
($10^{-5}{\rm eV^2}<\Delta m^2_{solar}<10^{-4}{\rm eV^2}$), 
the large angle 
low mass solution ($ \Delta m^2_{solar}\sim 10^{-7}{\rm eV^2}$)
and the Just-so solution 
($10^{-11}{\rm eV^2}<\Delta m^2_{solar}<10^{-10}{\rm eV^2}$). 
In order to discriminate these solutions, 
the Super-K made the extensive study on  
the flux independent analysis[6] by observing 
the day/night flux difference, the energy spectrum distortion 
of the recoil electron and the seasonal variation. 
Although the statistic is not sufficient, there is a tendency 
that the large mixing angle solutions are preferable. 
If the flux of the hep neutrino is taken free, the large 
angle MSW and the large angle low mass solutions have 
advantage[6]. These are signs to support that the solar neutrino 
calls for the large mixing between $\nu_e$ and $\nu_\mu$. 

At present, three typical mixing schemes to realize 
large mixing both for the atmospheric neutrino 
 and the solar neutrino mixings 
are known, 
the tri-maximal mixing[9], the bi-maximal mixing[10] and the 
democratic mixing[11]. Among them, the bi-maximal mixing 
and the democratic mixing matrix contain no CP violation phase. 
The reason is due to the absence of the mixing 
between the first and the third mass eigenstates. 
In contrast, the tri-maximal mixing predicts the maximal 
CP violation, 
which is the inevitable consequence of its structure. 

In view of the interest in the structure to give the large mixing 
between $\nu_\mu$ to $\nu_\tau$  and the maximal CP violation 
in the tri-maximal mixing that are derived from a democratic 
mass matrix as we see later,  we 
propose a democratic-type neutrino mass matrix based on $Z_3$ symmetry. 
We expected that this mass matrix interpolates 
the tri-maximal mixing scheme and the bi-maximal mixing scheme. 
Surprisingly,  we found that this mass matrix predicts that 
$\cos^2 \theta_{23}=\sin^2 \theta_{23}=1/2$. Here, we used 
$\theta_{ij}$ for the 
mixing angle between  mass eigenstates, $\nu_i$ and $\nu_j$. 
This relation is mostly needed to realize the large atmospheric 
neutrino mixing. 
We also found that this model predicts  
the CP violation phase, $\delta=\pi/2$. In our model, the 
mixing angle  between $\nu_1$ and $\nu_2$, $\theta_{12}$, and  
the mixing angle between  $\nu_1$ and $\nu_3$ , $\theta_{13}$, are 
left free. In order to examine the CP violation effect, 
we calculated the Jarlskog parameter and found that 
it takes about half of its maximal value if the large angle 
solar neutrino solutions are taken. 
 
In Sec.2, we give the democratic-type neutrino mass matrix. 
In Sec.3, the mixing matrix which is predicted by the mass 
matrix is derived and the physical implication is 
discussed.  The possible derivation of the democratic-type neutrino  
mass matrix is presented based on $Z_3$ symmetry in Sec.4. 
In Sec.5, the summary is given.

\section{Democratic-type neutrino mass matrix} 

Throughout of this paper, we consider the neutrino 
mass matrix in the diagonal mass basis of charged leptons. 
The name of the democratic-type for mass matrix is used so 
that the mass matrix includes the democratic forms of 
matrices and their deformations. 
  
\vskip 3mm
\noindent
(a) Democratic mass matrix

We first define the democratic forms of matrices 
which are the following matrices as 
\bea
S_{1}=\frac13\pmatrix{1&\z^2&\z\cr \z^2&\z&1\cr 
         \z&1&\z^2}\;,\;
S_{2}=\frac13\pmatrix{1&\z&\z^2\cr \z&\z^2&1\cr 
         \z^2&1&\z\cr }\;,
\;S_{3}=\frac13 \pmatrix{1&1&1\cr 1&1&1\cr 1&1&1\cr}\;,
\ena
where $\z=\exp(i2\pi/3)$ or $\exp(i4\pi/3)$ which satisfies $\z^3=1$
and $1+\z+\z^2=0$. The matrix 
$S_3$ is commonly referred to as a democratic form[11], 
but we consider the other two have the same right to be 
called democratic forms, because these matrices 
are related each other by the phase transformation as
\bea
PS_{1}P=S_{2}\;,\;PS_{2}P=S_{3}\;,\;PS_{3}P=S_{1}\;,
\ena
where
\bea
P=\pmatrix{1&0&0\cr 0&\z^2&0\cr 0&0&\z} \;,
\ena
and thus $S_1$ and $S_2$ are derived from $S_3$ by the phase 
transformation. 
It may be worthwhile to note that the phase matrix $P^*$ 
transforms $S_{i}$ in the reverse cyclic direction as 
$P^*S_{2}P^*=S_{1}$. 

We define the democratic mass matrix by the linear combination 
of these three democratic matrices as
\bea
m_{\nu,demo}=m_1^0 S_1+m_2^0 S_2+m_3^0 S_3\;. 
\ena
Here we consider that  mass parameters $|m_i^0|$ 
are quantities of the same order of magnitude, following 
the spirit of the democratic form. 

\vskip 3mm
\noindent
(b)  The deformation from the 
democratic mass matrix 

The deformation from the democratic form 
can be achieved by using the following three matrices,
\bea
T_1=\pmatrix{1&0&0\cr 0&\z&0\cr 0&0&\z^2\cr}\;,\;
T_2=\pmatrix{1&0&0\cr 0&\z^2&0\cr 0&0&\z\cr}\;,\;
T_3=\pmatrix{1&0&0\cr 0&1&0\cr 0&0&1\cr}\;.
\ena
Other symmetric mass matrices are formed by the linear 
combinations of $S_{i}$ and $T_i$. 
Thus, the general mass matrix is given by
\bea
m_\nu&=& m_{\nu,demo}+\tilde m_1 T_1
+\tilde m_2 T_2+\tilde m_3 T_3\nonumber\\
&=& \frac{1}{3}
\pmatrix{
\bar m_1+\bar m_2+\bar m_3& m_1^0\z^2+m_2^0\z+m_3^0&
 m_1^0\z+m_2^0\z^2+m_3^0\cr
 m_1^0\z^2+m_2^0\z+m_3^0 & \bar m_1\z+\bar m_2\z^2+\bar m_3&
 m_1^0+m_2^0+m_3^0\cr
 m_1^0\z+m_2^0\z^2+m_3^0& m_1^0+m_2^0+m_3^0&
 \bar m_1\z^2+\bar m_2\z+\bar m_3\cr}
\;,
\ena
where $\bar m_i=m_i^0 +3\tilde m_i $. In the following, 
we call $m^0_i$ (or $\bar m_i$) and $\tilde m_i$ 
mass parameters.   We call this mass matrix 
as the democratic-type mass matrix.

\section{Neutrino mixing matrix} 

The democratic-type mass matrix contains six complex 
parameters and thus it is a general matrix. 
 In order to reduce the degree of 
freedom, we assume that
 
"{\it all mass parameters, $m_i^0$ and 
$\tilde m_i$ are real"}. 
 
With this  assumption, the mass matrix contains 
six real freedoms which correspond 
to neutrino masses and mixing angles. Thus, in general 
the CP violation phases are predicted once neutrino masses 
and the mixing angles are given.

This assumption is one of the cases of the 
rather mild ansatz {\it "mass parameters are  proportional to 
either one of three quantities, 1, $\z$ and $\z^2$"}.  
Two other possibilities along this ansatz are discussed 
in Appendix B. 

In our model there are two cases, $\omega=e^{i2\pi/3}$ and 
$e^{i4\pi/3}$ which is the complex conjugate 
to $e^{i2\pi/3}$. The mass matrix $m_\nu$ with real mass 
parameters has the following property
\bea
m_\nu(\omega=e^{i2\pi/3})=m_\nu^*(\omega=e^{i4\pi/3})\;.
\ena
The neutrino mixing matrix $V$ is defined by $V^Tm_\nu V=D_\nu$ 
where $D_\nu={\rm diag} (m_1,m_2,m_3)$. If $V$ is the unitary 
matrix to diagonalize $m_\nu(\omega=e^{i2\pi/3})$, then 
$V^*$ is the one for $m_\nu(\omega=e^{i4\pi/3})$.  
In below, we discuss the neutrino mixing  matrix 
$V$  for $\omega=e^{i2\pi/3}$, by keeping in mind that 
$V^*$ is also allowed in our model. 

\vskip 3mm 
\noindent
(a) The neutrino mixing matrix 

We consider $m_{\nu}$ for $\omega=e^{i2\pi/3}$. We first transform  
mass matrix by using the tri-maximal mixing matrix $V_T$ as 
$V^T_{T} m_{\nu} V_T$, where
\bea
V_T=\frac{1}{\sqrt{3}}\pmatrix{
1 & 1 & 1 \cr
\z & \z^2 & 1 \cr
\z^2 & \z & 1
}\;.
\ena
Surprisingly, we find that the transformed mass matrix is 
a real symmetric matrix:
\bea
\tilde{m}_{\nu}=V^T_{T} m_{\nu} V_T 
=\pmatrix{
   m^0_1+\tilde{m}_1&\tilde m_3&\tilde m_2\cr
   \tilde m_3&m^0_2+\tilde{m}_2&\tilde m_1\cr
   \tilde m_2&\tilde m_1&m^0_3+\tilde{m}_3\cr}\;.
\ena
Then, the matrix  $\tilde m_\nu$ is diagonalized by an orthogonal 
matrix $O$.  

Now, the unitary matrix $V$ which diagonalizes $m_\nu$ is expressed 
by
\bea
V&=&\! V_T O\nonumber\\
 &=&\!\frac1{\sqrt 3} \!
    \pmatrix{O_{11}+O_{21}+O_{31}& O_{12}+O_{22}+O_{32}&
    O_{13}+O_{23}+O_{33}\cr
    \z O_{11}+\z^2 O_{21}+O_{31}& \z O_{12}+\z^2 O_{22}+O_{32}& 
    \z O_{13}+\z^2 O_{23}+O_{33}\cr
     \z^2 O_{11}+\z O_{21}+O_{31} & \z^2 O_{12}+\z O_{22}+O_{32} &
      \z^2 O_{13}+\z O_{23}+O_{33} \cr}\;.
      \nonumber\\
\ena
This unitary matrix is the neutrino mixing matrix because we 
consider the neutrino mass matrix in the diagonal mass basis 
of charged leptons. This mixing matrix seems to have a complex 
form, but it has an outstanding property that 
$V_{2i}=V_{3i}^*$ for $i=1,2,3$. This property restricts the neutrino 
mixings tightly. Since it is hard to treat this mixing matrix 
directly, we attack it from slightly different point of view. 

We first observe that by the phase transformation of 
charged leptons and neutrinos,  
$V$ can be made into the standard form 
$V_{SF}$ as given in the particle data[12]. 
\bea
V_{SF}= 
 \pmatrix{c_{12}c_{13}&s_{12}c_{13}&s_{13}e^{-i\delta}\cr
 -s_{12}c_{23}-c_{12}s_{23}s_{13}e^{i\delta}&
  c_{12}c_{23}-s_{12}s_{23}s_{13}e^{i\delta}& s_{23}c_{13}
  \cr
  s_{12}s_{23}-c_{12}c_{23}s_{13}e^{i\delta}&
  -c_{12}s_{23}-s_{12}c_{23}s_{13}e^{i\delta}& c_{23}c_{13}
  \cr}\;,
\ena
where $s_{ij}=\sin \theta_{ij}$, $c_{ij}=\cos \theta_{ij}$ 
and $\theta_{ij}$ is the mixing angle which mixes 
mass eigenstates $\nu_i$ and $\nu_j$. 
That is, we can write $V=PV_{SF}P'$, where 
$P$ and $P'$ are diagonal phase matrices. 

The restrictions $V_{2i}=V_{3i}^*$ for $i=1,2,3$ 
lead to the constraints    
$|(V_{SF})_{2i}|=|(V_{SF})_{3i}|$ for $i=1,2,3$, 
which are expressed by 
\bea
|-s_{12}c_{23}-c_{12}s_{23}s_{13}e^{i\delta}|
&=&|s_{12}s_{23}-c_{12}c_{23}s_{13}e^{i\delta}|\;,\nonumber\\
|c_{12}c_{23}-s_{12}s_{23}s_{13}e^{i\delta}|
&=&|-c_{12}s_{23}-s_{12}c_{23}s_{13}e^{i\delta}|\;,\nonumber\\
|s_{23}c_{13}|&=&|c_{23}c_{13}|\;.
\ena
By solving these equations, we find
\bea
c_{23}^2=s_{23}^2\;,\;\quad  \cos \delta=0\;,
\ena
by omitting the uninteresting possibility $c_{13}=0$. 
It is amazing that our model predicts the  CP violation phase, 
$\delta=\pi/2$, 
and $c_{23}^2=s_{23}^2=1/2$ which is quite important to explain 
the almost full mixing between $\nu_\mu$ and $\nu_\tau$ 
in the two mixing limit. The most interesting point is 
that the mixing angle $\theta_{23}$ and the CP violation 
phase $\delta$ are fixed independently of mass parameters. 

\vskip 3mm
\noindent
(b) General form of neutrino mixing matrix

We take $s_{23}=-c_{23}=-\frac {1}{\sqrt 2}$. Then, the 
diagonal phase matrices $P$ and $P'$ are determined 
such that the matrix $V_T^\dagger PV_{SF}P'$ becomes 
a real orthogonal matrix. In this way, we found  
\bea
V=\pmatrix{1&0&0\cr0&e^{i\rho}&0\cr0&0&e^{-i\rho}\cr}
 \pmatrix{c_{12}c_{13}&s_{12}c_{13}&-is_{13}\cr
 -\frac{s_{12}-ic_{12}s_{13}}{\sqrt 2}&
 \frac{c_{12}+is_{12}s_{13}}{\sqrt 2}&-\frac{c_{13}}{\sqrt 2}\cr
 -\frac{s_{12}+ic_{12}s_{13}}{\sqrt 2}&
 \frac{c_{12}-is_{12}s_{13}}{\sqrt 2}&\frac{c_{13}}{\sqrt 2}\cr}
 \pmatrix{1&0&0\cr0&1&0\cr0&0&i\cr}\;.
\ena

In addition to interesting predictions for $\theta_{23}$ and 
$\delta$, our neutrino mass matrix predicts the Majorana 
phase matrix ${\rm diag}(1,1,i)$[13, 14] which shows no 
CP violation intrinsic to Majorana system. 
The other phase matrix 
${\rm diag}(1,e^{i\rho},e^{-i\rho})$ does not have any 
physical effect, because this phase is 
absorbed by charged leptons.    
Our mass matrix contains six real parameters which are 
converted to three neutrino masses, two mixing angles, 
$\theta_{12}$ and $\theta_{13}$, and one unphysical phase 
$\rho$. 

The other case $s_{23}=c_{23}=\frac {1}{\sqrt 2}$ reduces to 
the case of $\delta=-\pi/2$, which is included in the mixing 
matrix $V^*$. 

In below, we discuss that our mixing reduces to 
two well-known typical large mixing matrices, 
the tri-maximal mixing and the bi-maximal mixing 
by imposing simple conditions on mass parameters. 

\noindent
(c) Tri-maximal and Bi-maximal mixing limits

By taking the mass parameters in some special values, 
our model reduces to models to reproduce the tri-maximal 
mixing and the bi-maximal mixing.

\noindent
(c-1) The tri-maximal mixing limit

By taking the mixing angles and phase matrices as 
$s_{12}=-1/\sqrt{2}$, $c_{12}=1/\sqrt{2}$, 
$s_{13}=1/\sqrt{3}$, $c_{13}=\sqrt{2/3}$, 
$\rho=\pi/2 $, the matrix $V$ reduces to the 
tri-maximal mixing matrix 
\bea
V=V_T\pmatrix{1&0&0\cr 0&-1&0\cr 0&0&1\cr}\;,
\ena
where the phase matrix ${\rm diag}(1,-1,1)$ does not 
have any physical meaning. 

From Eq.(A.2) in Appendix, the mass parameters are 
now restricted by 
\bea
&&m_1^0=m_1\;,\;m_2^0=m_2\;,\;m_3^0=m_3\;, 
\nonumber\\
&&\tilde{m}_1=\tilde m_2=\tilde m_3=0\;.
\ena
Now we see the mass matrix $m_{\nu}$ which is  
 reduced to the $m_{\nu,demo}$ as 
\bea
m_\nu=m_1^0 S_1+m_2^0 S_2+m_3^0 S_3\;.
\ena
The democratic mass matrix $m_{\nu, demo}$ has 
various interesting properties which are discussed in 
Appendix A.

\noindent
(c-2) The bi-maximal mixing limit

By taking the mixing angles and phase matrices as 
$s_{12}=-1/\sqrt{2}$, $c_{12}=1/\sqrt{2}$, 
$s_{13}=0$, $c_{13}=1$, 
$\rho=0 $, the matrix $V$ reduces to the 
bi-maximal mixing matrix 
\bea
V=\pmatrix{1&0&0\cr 0&1&0\cr 0&0&1\cr}
O_B \pmatrix{1&0&0\cr 0&1&0\cr 0&0&i\cr},
\ena
where $O_B$ is the bi-maximal mixing matrix defined 
by 
\bea
O_B=\pmatrix{
\frac{1}{\sqrt{2}} & -\frac{1}{\sqrt{2}} & 0 \cr
\frac{1}{2} & \frac{1}{2} & -\frac{1}{\sqrt{2}} \cr
\frac{1}{2} & \frac{1}{2} & \frac{1}{\sqrt{2}}
}\;.
\ena

From Eq.(A.2) in Appendix, the mass parameters are 
now restricted by 
\bea
m_1^0=m_2^0\;,\; \tilde m_1=\tilde m_2\;,
\ena
and in this case the mass matrix becomes 
\bea
m_{\nu,B}=\frac{1}{3}\pmatrix{
\bar m_3+2\bar m_1 & m_3^0-m_1^0&m_3^0-m_1^0\cr
  m_3^0-m_1^0 & \bar m_3 -\bar m_1&
 m_3^0+2m_1^0 \cr
 m_3^0-m_1^0 & m_3^0+2 m_1^0 &
 \bar m_3 -\bar m_1\cr}
\;.
\ena
This matrix satisfies the condition that all elements are real, 
$(m_{\nu,B})_{22}= (m_{\nu,B})_{33}$ and 
$(m_{\nu,B})_{12}= (m_{\nu,B})_{13}$. 
  
The mass parameters are expressed by neutrino masses and 
mixing angles as
\bea
m_1^0=m_2^0&=&\frac14 (2m_3+m_2+m_1)+
       \frac{1}{2\sqrt 2}(m_2-m_1)\;,\nonumber\\
m_3^0&=&\frac14 (2m_3+m_2+m_1)-
       \frac{1}{\sqrt 2}(m_2-m_1)\;,\nonumber\\
\tilde m_1=\tilde m_2&=&-\frac{1}{6\sqrt 2}(m_2-m_1)\;,
\nonumber\\
\tilde m_3&=&-\frac14(2m_3-m_2-m_1)+\frac{1}{3\sqrt 2}(m_2-m_1)
\;.
\ena

It is interesting to observe that our model connects 
the tri-maximal mixing and the bi-maximal mixing 
by keeping the CP violation phase, $\delta=\pi/2$. 
In our model, the absence of the CP violation in the 
bi-maximal limit is solely due to $\sin \theta_{13}=0$ 
and any deviation from it recovers $\delta=\pi/2$. 
Since the restriction 
$\sin^2 \theta_{23}=\cos^2 \theta_{23}=1/2$ is 
the most advantageous situation to realize large 
mixing angle $\sin^2 2\theta_{atm}$ by deviating 
$\sin \theta_{13}$ from zero, this model provides 
the most advantageous case for the CP violation.

\section{Analysis of our mixing scheme}

We consider  the hierarchy of neutrino masses  as 
\bea
\Delta_{atm}\equiv \Delta_{32}& \simeq& \Delta_{31}
\simeq 3\times 10^{-3}{\rm eV}^2\;,
\nonumber \\
\Delta_{solar}\equiv \Delta_{21}& << & \Delta_{atm} \;.
\ena
 
\vskip 3mm
\noindent
(a) Vacuum oscillations 

We first derive the probabilities of neutrino oscillations 
in the vacuum. We use the abbreviation,
$P(\ell \to \ell')$ for $P(\nu_\ell \to \nu_{\ell'})$. We find 
\bea
P(\tau \to \tau)=P(\mu \to \mu)\;,\;
P(e \to \tau)=P(\mu \to e)\;,\;
P(\tau \to e) =P(e \to \mu)
\ena
and
\bea
P(e\to e)&=& 1-4s_{12}^2c_{12}^2c_{13}^4
      \sin^2\left(\frac{\Delta_{21}}{4E}L\right)
      -4c_{12}^2 s_{13}^2c_{13}^2
 \sin^2\left(\frac{\Delta_{31}}{4E}L\right)
      \nonumber\\
&&\hskip 10mm -4s_{12}^2 s_{13}^2c_{13}^2
 \sin^2\left(\frac{\Delta_{32}}{4E}L\right)\;,
  \nonumber\\
P(\mu \to \mu)&=&1- A^2B^2\sin^2\left(\frac{\Delta_{21}}{4E}L
            \right)-c_{13}^2A^2
          \sin^2\left(\frac{\Delta_{31}}{4E}L\right)
     \nonumber\\
    && \hskip 10mm -c_{13}^2B^2
     \sin^2\left(\frac{\Delta_{32}}{4E}L\right)\;,
     \nonumber\\
P(e \to \mu)&=&
    \frac12 c_{13}^2\left [s_{13}+c_{12}A +s_{12}B \right]^2
    -2s_{12}c_{12}c_{13}^2 AB\sin^2\left(
     \frac{\Delta_{21}}{4E}L+
      \frac{\delta_{1}}{2}+\frac{\delta_2}{2}\right)
    \nonumber\\
    &&-2 c_{12}s_{13}c_{13}^2 A\sin^2\left(
     \frac{\Delta_{31}}{4E}L +\frac{\delta_{1}}2-\frac{\pi}{4}
      \right)
    -2s_{12}s_{13}c_{13}^2B 
    \sin^2\left(
    \frac{\Delta_{32}}{4E}L- \frac{\delta_{2}}2-\frac{\pi}{4}\right)\;, 
    \nonumber\\
 P(\mu \to e)&=&\frac12 c_{13}^2\left [s_{13}+c_{12}A +
      s_{12}B \right]^2
     -2s_{12}c_{12}c_{13}^2 AB\sin^2\left(
     \frac{\Delta_{21}}{4E}L-
     \frac{\delta_{1}}{2}-\frac{\delta_2}{2}\right)
   \nonumber\\
     &&-2 c_{12}s_{13}c_{13}^2 A\sin^2\left(
    \frac{\Delta_{31}}{4E}L-\frac{\delta_{1}}2+\frac{\pi}{4}
    \right)
     -2s_{12}s_{13}c_{13}^2B \sin^2\left(
   \frac{\Delta_{32}}{4E}L+\frac{\delta_{2}}2+\frac{\pi}{4}\right)\;, 
  \nonumber\\
P(\mu \to \tau)&=&1- 
  A^2B^2 \sin^2\left(
  \frac{\Delta_{21}}{4E}L-\delta_1-\delta_2\right)
  -c_{13}^2 A^2\sin^2\left(
  \frac{\Delta_{31}}{4E}L- \delta_{1}-\frac{\pi}2 \right)
 \nonumber\\
 &&\hskip 10mm -c_{13}^2 B^2
 \sin^2\left(
 \frac{\Delta_{32}}{4E}L+ \delta_{2}-\frac{\pi}2 \right)\;,
 \nonumber\\
P(\tau \to \mu)&=&1- A^2B^2 \sin^2\left(
  \frac{\Delta_{21}}{4E}L+ \delta_1+\delta_2\right)
  -c_{13}^2 A^2\sin^2\left(
 \frac{\Delta_{31}}{4E}L + \delta_{1}-\frac{\pi}2 \right)
 \nonumber\\
 &&\hskip 10mm -c_{13}^2 B^2\sin^2\left(
 \frac{\Delta_{32}}{4E}L- \delta_{2}-\frac{\pi}2 \right)\;,
\ena 
where
\bea
A&=&\sqrt{s_{12}^2+c_{12}^2s_{13}^2}\;,\;
B=\sqrt{c_{12}^2+s_{12}^2s_{13}^2}\;,\nonumber\\
\delta_1&=&\tan^{-1}\left(\frac{c_{12}s_{13}}{s_{12}}\right)\;,\;
\delta_2=\tan^{-1}\left(\frac{s_{12}s_{13}}{c_{12}}\right)\;,
\ena
and $\Delta_{ij}=m_i^2-m_j^2$.  These are general formula and 
the simpler form of the oscillation probability is obtained 
once the distance $L$ is specified.

\vskip 3mm
\noindent
(b) The analysis

We start from the CHOOZ data which restrict 
$|V_{e3}|^2<0.05$ which leads to 
\bea
s_{13}^2<0.05\;.
\ena
Next, the probability of $\nu_\mu$ to $\nu_e$ and $\nu_\tau$ 
at the atmospheric range are simply expressed by 
\bea
P(\mu \to e)
&\simeq& 2s^2_{13}c^2_{13}\sin^2{\frac{ \Delta_{atm} }{4E} L }\;,
\nonumber\\
P(\mu \to \tau)
&\simeq& c^4_{13}\sin^2{\frac{ \Delta_{atm} }{4E} L }\;.
\ena
Therefore, by combining our model and the CHOOZ data we 
predict the probability for $\nu_\mu$ to $\nu_e$ is small 
, $ P(\mu \to e)<0.1$ 
and the effective mixing angle between $\nu_\mu$ to $\nu_\tau$ is
\bea
\sin^2 2\theta_{atm}=c_{13}^4> 0.90\;.
\ena

As for the solar neutrino problem, we assume  
$10^{-11}{\rm eV^2} <\Delta_{solar} < 10^{-4}{\rm eV^2}$. 
In the vacuum, we find 
\bea
P(\nu_e \rightarrow \nu_e)
&\simeq& 1-2s^2_{13}c^2_{13}-
          4s^2_{12}c^2_{12}c^4_{13}\sin^2{\frac{ \Delta_{solar} }{4E}
L
}\;,
\nonumber\\
P(\nu_e \rightarrow \nu_\mu)
&\simeq& s^2_{13}c^2_{13} + 
          2s^2_{12}c^2_{12}c^4_{13}\sin^2{\frac{ \Delta_{solar} }
          {4E} L}+
          s_{12}c_{12}s_{13}c^2_{13}\sin{\frac{ \Delta_{solar} }
          {2E} L
}\;,
\nonumber\\
P(\nu_e \rightarrow \nu_\tau)
&\simeq& s^2_{13}c^2_{13} + 
          2s^2_{12}c^2_{12}c^4_{13}\sin^2{\frac{ \Delta_{solar} }
          {4E} L}-
          s_{12}c_{12}s_{13}c^2_{13}\sin{\frac{ \Delta_{solar} }
          {2E} L
}\;.
\ena 
Thus, we find that 
\bea
\sin^2 2\theta_{solar}\simeq \sin^2 2\theta_{12}c_{13}^4
>0.90 \sin^2 2\theta_{12}\;.
\ena
Thus, our model can accommodate all four solutions,  
the small angle MSW, the 
large angle MSW, the low mass and the Just-so solutions. 

\vskip 3mm
\noindent
(c) CP violation 
 
In order to see the size of the CP violation, we 
consider the Jarlskog parameter that is defined by[15]
\bea
J_{CP}\equiv \Im (V_{e1}V_{e2}^*V_{\mu 1}^*V_{\mu 2})=
s_{12}c_{12}s_{23}c_{23}s_{13}c_{13}^2\sin \delta
\le 1/6\sqrt{3} \;.
\ena

Our predicted values, $c_{23}=-s_{23}=1/\sqrt2$ and 
$\sin \delta=1$  
give the most advantageous case to obtain large 
$J_{CP}$ concerning $\theta_{23}$ and $\delta$,
\bea
(J_{CP})_{our\; model}=-\frac12 s_{12}c_{12}
s_{13}c_{13}^2 \;.
\ena
The prediction of $J_{CP}$ depends on $\theta_{12}$ and 
$\theta_{13}$. If we take the value $s_{13}^2=0.05$, we 
have $J_{CP}=-0.053\sin 2\theta_{12}$. If the solar neutrino 
mixing turns out to be one of large angle solutions, 
$ \sin^2 2\theta_{12}\sim 0.8$ we find  $J_{CP}=-0.047$ 
which is about half of the maximal
value $(J_{CP})_{max}\simeq 0.096$. For the small angle case, 
we obtain about 10 times smaller value than the large angle case.

\section{Some derivations of the neutrino mass matrix}
 
The neutrino mass matrix that we discussed in the 
former section may be derived by following considerations 
in the basis where charged leptons are mass eigenstates. 

\noindent
(a) Neutrino mass term and $S_3$ symmetry with $Z_3$ phases

We consider the following three types of transformations;
\bea
{\rm (I)}&& \nu_e \to \z^2 \nu_\mu\;,\;\nu_\mu \to \z^2 \nu_\tau\;,\;
\nu_\tau \to \z^2 \nu_e\;,\nonumber\\
{\rm (II)}&& \nu_e \to \z \nu_\mu\;,\;\nu_\mu \to \z \nu_\tau\;,\;
\nu_\tau \to \z \nu_e\;,\nonumber\\
{\rm (III)}&& \nu_e \to  \nu_\mu\;,\;\nu_\mu \to \nu_\tau\;,\;
\nu_\tau \to \nu_e\;,\;
\ena
where  $\z=\exp(i2\pi/3)$ or $\exp(i4\pi/3)$. They are considered 
as $S_3$ transformations with $Z_3$ phases. 

The  Majorana mass matrix for left-handed neutrinos 
which is invariant under one of these transformation 
is expressed by
\bea
M_i={m_i^0} S_i+ \tilde {m_i} T_i\;,
\ena
where $i=1,2,3$, $S_i$ and $T_i$ are defined in Eqs.(2) and 
(6). The mass matrix $M_1$ is derived by imposing the 
transformation (I) and so on. 
  
Since there is no principle to discriminate these three 
matrices $M_i$,  we assume that the neutrino mass matrix $m_\nu$ is 
expressed by the sum of  these three mass matrices, although there 
is no good reason to explain this. Then, 
we obtain the neutrino mass matrix, $m_\nu$ in Eq.(7). 

\noindent
(b) $Z_3$ invariant Lagrangian 

Another reason to introduce the mass matrix in Eq.(7) may be 
given by imposing  
the $Z_3$ symmetry on Yukawa interaction. 
The left-handed doublet leptons can be arranged in 
eigenstates of $Z_3$ symmetry as 
\bea
\Psi_1&=&\frac{\z^2 \ell_e+\z \ell_\mu+\ell_\tau}{\sqrt 3}\;,
\nonumber\\
\Psi_2&=&\frac{\z \ell_e+\z^2 \ell_\mu+\ell_\tau}{\sqrt 3}\;,
\nonumber\\
\Psi_3&=&\frac{\ell_e+\ell_\mu+\ell_\tau}{\sqrt 3}\;,
\ena
where $\ell_e^T=(\nu_{eL},e_L)$ and so on. Under the 
$S_3$ transformation, $\ell_e\to \ell_\mu$ and 
$\ell_\mu\to \ell_\tau$ $\ell_\tau\to \ell_e$, they are  
transformed as 
\bea
\Psi_1\to \z^2 \Psi_1\;,\;
\Psi_2\to \z \Psi_2\;,\; \Psi_3\to \Psi_3\;.
\ena

Then, we introduce three kinds of triplet Higgs which 
transform as $\Delta_1 \to \z^2 \Delta_1$,  
$\Delta_2 \to \z \Delta_2$ and $\Delta_3 \to \Delta_3$. 
Then, the invariant Yukawa interaction terms 
among two doublets and a triplet are  
\bea
{\cal L}_{\rm y}&=&-\Biggl(
 {(m^0_1+\tilde{m}_1)\z^2} \overline{(\Psi_1)^C}i\tau_2 
 \frac{\Delta_1}{v_1} \Psi_1
 +{ (m^0_2+\tilde{m}_2) \z}\overline{(\Psi_2)^C}i\tau_2
 \frac{\Delta_2}{v_2}\Psi_2\nonumber\\
 &&\mbox{ }+(m^0_3+\tilde{m}_3) \overline{(\Psi_3)^C}i\tau_2
 \frac{\Delta_3}{v_3}\Psi_3 \Biggr)
 \nonumber\\
 &&-2 \left(\tilde m_1 \z^2 \overline{(\Psi_2)^C}i\tau_2
 \frac{\Delta_1}{v_1} \Psi_3
 +\tilde m_2 \z \overline{(\Psi_3)^C}i\tau_2
 \frac{\Delta_2}{v_2} \Psi_1
 +\tilde m_3 \overline{(\Psi_1)^C}i\tau_2
 \frac{\Delta_3}{v_3} \Psi_2\right)
 \;,
\ena
where $v_i$ are vacuum expectation values of $\Delta_i$. 
When vacuum expectation values of triplet Higgs are generated, 
the Majorana-type mass term given in Eq.(7) 
is generated for neutrinos.  
We argue that in order to acquire small vacuum expectation values 
of triplet Higgs bosons, the seesaw suppression mechanism[16] 
should be adopted. 

\noindent
(c) Non-renormalizable interaction

The triplet representation can be composed of two doublet 
representation. We can explicitly construct the  Higgs triplet, 
$\Delta_i$ by the combinations of two  Higgs doublets, $H_j$ 
which  transform as  
\bea
H_1 \rightarrow \z H_1\;,\; H_2 \rightarrow H_2\;.
\ena
The symmetric combinations $H_1H_1$, $H_1H_2$ and $H_2H_2$ 
transform as $\Delta_1$, $\Delta_2$ and $\Delta_3$. 
Thus, we obtain the  Lagrangian as 
\bea
{\cal L}_{\rm y}
&=&-
  \Biggl(
     (m^0_1+\tilde{m}_1)\z^2\overline{(\Psi_1)^C} \Psi_1 
                        \frac{H_1 H_1}{u^2_1}+
     (m^0_2+\tilde{m}_2)\z\overline{(\Psi_2)^C} \Psi_2 
                        \frac{H_1 H_2}{u_1u_2}  \nonumber\\
     &&\mbox{ }
      +(m^0_3+\tilde{m}_3)\overline{(\Psi_3)^C} \Psi_3 
                        \frac{H_2 H_2}{u^2_2} \Biggr)\nonumber\\
&&-2\left(
     \tilde{m}_1\z^2\overline{(\Psi_2)^C} \Psi_3 \frac{H_1
H_1}{u^2_1}+
     \tilde{m}_2\z\overline{(\Psi_1)^C} \Psi_3 \frac{H_1 H_2}{u_1u_2}+
     \tilde{m}_3\overline{(\Psi_1)^C} \Psi_2 \frac{H_2 H_2}{u^2_2}  
  \right)\;,
\ena
where $u_i$ is the vacuum expectation value of the neutral 
component of $H_i$. After the symmetry breaking, the neutrino 
mass matrix in Eq.(7) is obtained.

\section{Discussions}
 
We introduced the democratic-type neutrino mass matrix by 
extending the democratic mass matrix and 
found that one angle $\theta_{23}$ and the CP violation 
phase intrinsic to the Dirac system are predicted to be 
$\theta_{23}=-\pi/4$ and $\delta =\pi/2$. As a consequence, 
the mixing matrix is expressed by two angles, 
$\theta_{12}$ and $\theta_{13}$, as shown in Eq.(15).   
If the solar neutrino problem turns out to be solved by 
the large angle solutions,  the large CP violation effect 
is expected. In this situation, our model predicts 
that the Jarlskog parameter is about half of the 
maximal value, $J_{CP}=-0.047$ with 
$\sin^2 2\theta_{solar}=0.8$. This could be 
explored by the future long-baseline experiments.
 
Our model predicts no CP violation intrinsic to 
Majorana neutrino system[13],[14]. The phase $i$ in the 
Majorana phase matrix ${\rm diag}(1,1,i)$ 
in Eq.(15) relates to the CP signs 
of mass eigenstate neutrinos[17] in addition to relative 
signs of neutrino masses. 
The phase matrix  ${\rm diag}(1,e^{i\rho},e^{-i\rho})$ in Eq.(15) 
are absorbed by charged leptons.  

The effect for the neutrinoless double beta decay 
 is given by[18]
\bea
|<m_\nu>|\equiv |{\sum_j}' U_{ej}^2 m_j|
=|(m_1c_{12}^2+m_2s_{12}^2)c_{13}^2+m_3s_{13}^2|\;, 
\ena
where the dash in the sum 
means that $j$ extends to light neutrinos. The mixing 
matrix  $U$ is the matrix including the Majorana phase matrix, 
$U=V_{SF}{\rm diag}(1,1,i)$. 
The effective mass $|<m_\nu>|$ 
depends on 
the relative signs among $m_1$, $m_2$ and $m_3$, which corresponds 
to CP signs of mass eigenstate neutrinos[17]. Here we take 
$m_1>0$.  In  case that $|m_1|\simeq |m_2|$, we find 
\bea
|<m_\nu>| 
=\cases{|m_1c_{13}^2+m_3s_{13}^2| \qquad\qquad\quad\> (m_2>0)\cr
        |m_1\cos 2\theta_{12}c_{13}^2+m_3s_{13}^2|\qquad\! (m_2<0)\cr}\;.
\ena
There are three typical cases. 
\begin{enumerate}
\item The similar mass case $|m_2|\sim |m_3|\sim m_1$ 

In this case, $\Delta_{solar}$ and $\Delta_{atm}$ do not 
constrain neutrino mass themselves. The effective mass 
$|<m_\nu>|\sim m_1$ or $m_1|\cos 2\theta_{12}|$ could be 
as large as the sensitivity of the neutrinoless double beta decay 
experiment. It may be worthwhile to comment that 
$\cos 2\theta_{12}\sim 1$ for the small angle solution 
and $\sim 0.44$ for the large angle solutions such as 
$\sin^2 2\theta_{solar}\simeq 0.8$ for the solar neutrino 
problem. 

\item The hierarchical case
\begin{enumerate}
\item $|m_3|>>m_1 \simeq |m_2|$ 

We expect that $|m_3|\sim \sqrt{\Delta_{atm}}\sim 0.05$eV. 
Then, we expect  $|<m_\nu>|<< |m_3|\sim 0.05$eV, which may 
be hard to be detected.

\item $m_1 \simeq |m_2|>>|m_3|$

We expect that $m_1\sim \sqrt{\Delta_{atm}}\sim 0.05$eV. 
Then, we expect  $|<m_\nu>|\sim m_1$ or $|\cos 2 \theta_{12}|m_1$ 
which is about the order  0.05eV, which may 
be within the reach of the future experiment.
\end{enumerate}
\end{enumerate}

\vskip .5mm

{\Huge Acknowledgment} 
This work is supported in part by 
the Japanese Grant-in-Aid for Scientific Research of
Ministry of Education, Science, Sports and Culture, 
No.11127209.

\vskip 1mm
Note added: After submitting our paper, we are informed 
by H. Fritzsch and Z.Z. Xing[19] that they discussed 
another possibility to obtain the large CP violation. 
\newpage

\setcounter{section}{0}
\renewcommand{\thesection}{\Alph{section}}
\renewcommand{\theequation}{\thesection .\arabic{equation}}
\newcommand{\apsc}[1]{\stepcounter{section}\noindent
\setcounter{equation}{0}{\Large{\bf{Appendix\,\thesection:\,{#1}}}}}

\apsc{Explicit expressions of mass parameters and the interesting 
property of the democratic mass matrix}

\vskip 3mm
\noindent
(a) Mass parameters 

Mass parameters $m_i^0$ and $\bar m_i=m_i^0+3\tilde m_i$ 
are explicitly expressed in 
terms of neutrino masses and mixing angles, $\theta_{12}$ and 
$\theta_{13}$, and the unphysical phase $\rho$ which 
is eaten by the phase redefinition of charged leptons. 
This is achieved by examining 
\bea
m_\nu=V^*\pmatrix{m_1&0&0\cr 0&m_2&0\cr 0&0&m_3\cr}V^\dagger\;.
\ena
We find for $\omega=e^{i2\pi/3}$
\bea
m^0_3&=&\left\{\frac{1}{2}(s^2_{12}+c^2_{12}s^2_{13})-\sqrt{2}
c_{12}c_{13}
(s_{12}\cos{\rho}+c_{12}s_{13}\sin{\rho})\right\}m_1
\nonumber\\
&&+\left\{\frac{1}{2}(c^2_{12}+s^2_{12}s^2_{13})+\sqrt{2}s_{12}c_{13}
(c_{12}\cos{\rho}-s_{12}s_{13}\sin{\rho})\right\}m_2
\nonumber\\
&&+\left\{\frac{1}{2}c^2_{13}+\sqrt{2}s_{13}c_{13}
\sin{\rho}\right\}m_3\;,
\nonumber\\
m^0_2&=&\left\{\frac{1}{2}(s^2_{12}+c^2_{12}s^2_{13})-\sqrt{2}
c_{12}c_{13}
\biggl(s_{12}\cos{\Bigl(\rho+\frac{2\pi}{3}\Bigr)}+c_{12}s_{13}
\sin{\Bigl(\rho+\frac{2\pi}{3}\Bigr)}\biggr)\right\}m_1
\nonumber\\
&&+\left\{\frac{1}{2}(c^2_{12}+s^2_{12}s^2_{13})+\sqrt{2}
s_{12}c_{13}
\biggl(c_{12}\cos{\Bigl(\rho+\frac{2\pi}{3}\Bigr)}-s_{12}s_{13}
\sin{\Bigl(\rho+\frac{2\pi}{3}\Bigr)}\biggr)\right\}m_2
\nonumber\\
&&+\left\{\frac{1}{2}c^2_{13}+\sqrt{2}s_{13}c_{13}
\sin{\Bigl(\rho+\frac{2\pi}{3}\Bigr)}\right\}m_3\;,
\nonumber\\
m^0_1&=&\left\{\frac{1}{2}(s^2_{12}+c^2_{12}s^2_{13})-\sqrt{2}
c_{12}c_{13}
\biggl(s_{12}\cos{\Bigl(\rho-\frac{2\pi}{3}\Bigr)}+c_{12}s_{13}
\sin{\Bigl(\rho-\frac{2\pi}{3}\Bigr)}\biggr)\right\}m_1
\nonumber\\
&&+\left\{\frac{1}{2}(c^2_{12}+s^2_{12}s^2_{13})+\sqrt{2}
s_{12}c_{13}
\biggl(c_{12}\cos{\Bigl(\rho-\frac{2\pi}{3}\Bigr)}-s_{12}s_{13}
\sin{\Bigl(\rho-\frac{2\pi}{3}\Bigr)}\biggr)\right\}m_2
\nonumber\\
&&+\left\{\frac{1}{2}c^2_{13}+\sqrt{2}s_{13}c_{13}
\sin{\Bigl(\rho-\frac{2\pi}{3}\Bigr)}\right\}m_3\;,
\nonumber\\
\bar{m}_3&=&\left\{c^2_{12}c^2_{13}+(s^2_{12}-c^2_{12}s^2_{13})
\cos{2\rho}
+2s_{12}c_{12}s_{13}\sin{2\rho}\right\}m_1
\nonumber\\
&&+\left\{s^2_{12}c^2_{13}+(c^2_{12}-s^2_{12}s^2_{13})\cos{2\rho}
-2s_{12}c_{12}s_{13}\sin{2\rho}\right\}m_2
\nonumber\\
&&+\left\{s^2_{13}-c^2_{13}\cos{2\rho}\right\}m_3\;,
\nonumber\\
\bar{m}_2&=&\left\{c^2_{12}c^2_{13}+(s^2_{12}-c^2_{12}s^2_{13})
\cos{\Bigl(2\rho-\frac{2\pi}{3}\Bigr)}
+2s_{12}c_{12}s_{13}\sin{\Bigl(2\rho-\frac{2\pi}{3}\Bigr)}\right\}m_1
\nonumber\\
&&+\left\{s^2_{12}c^2_{13}+(c^2_{12}-s^2_{12}s^2_{13})
\cos{\Bigl(2\rho-\frac{2\pi}{3}\Bigr)}
-2s_{12}c_{12}s_{13}\sin{\Bigl(2\rho-\frac{2\pi}{3}\Bigr)}\right\}m_2
\nonumber\\
&&+\left\{s^2_{13}-c^2_{13}\cos{\Bigl(2\rho-\frac{2\pi}{3}\Bigr)}
\right\}m_3\;,
\nonumber\\
\bar{m}_1&=&\left\{c^2_{12}c^2_{13}+(s^2_{12}-c^2_{12}s^2_{13})
\cos{\Bigl(2\rho+\frac{2\pi}{3}\Bigr)}
+2s_{12}c_{12}s_{13}\sin{\Bigl(2\rho+\frac{2\pi}{3}\Bigr)}\right\}m_1
\nonumber\\
&&+\left\{s^2_{12}c^2_{13}+(c^2_{12}-s^2_{12}s^2_{13})
\cos{\Bigl(2\rho+\frac{2\pi}{3}\Bigr)}
-2s_{12}c_{12}s_{13}\sin{\Bigl(2\rho+\frac{2\pi}{3}\Bigr)}\right\}m_2
\nonumber\\
&&+\left\{s^2_{13}-c^2_{13}\cos{\Bigl(2\rho+\frac{2\pi}{3}\Bigr)}
\right\}m_3 \;.
\nonumber\\
\ena

\noindent
(b) Interesting properties of the democratic mass matrix 
 
The democratic mass matrix $m_{\nu,demo}$ defined in Eq.(5) consists of  
matrices $S_i$ which are rank 1 and 
have a special property that 
they are diagonalized simultaneously by the 
bilinear transformation 
$V_T^TS_iV_T$ with the unitary matrix $V_T$. The condition 
that symmetric matrices $A$ and $B$ are diagonalized simultaneously 
by this  transformation is 
$A^* B=B^* A$ and matrices $S_i$ satisfy $S_i^*S_j=0$ for 
$i\neq j$, so that they satisfy the condition trivially. 
In fact, we find 
\bea
V_T^TS_i V_T=D_i\;,
\ena
where $D_{i}$ are diagonal matrix as 
$D_1={\rm diag}(1,0,0)$, $D_2={\rm diag}(0,1,0)$, 
$D_3={\rm diag}(0,0,1)$ and   
\bea
V_T=\frac{1}{\sqrt 3} 
\pmatrix{1&1&1\cr \z&\z^2&1\cr \z^2&\z&1\cr}\;.
\ena
 
By using $V_T$, the democratic neutrino mass matrix 
$m_{\nu,demo}$ in Eq.(5) is diagonalized 
as
\bea
V_T^T m_{\nu,demo} V_T=
\pmatrix{m_1^0&0&0\cr 0&m_2^0&0\cr 0&0&m_3^0\cr}\;.
\ena
Thus, in this limit $m_i^0$ are interpreted to be 
masses of neutrinos. 

The unitary matrix $V_T$ is nothing but the tri-maximal 
mixing matrix. 
The matrix $V_T$ is transformed into the standard form as
\bea
\pmatrix{1&0&0\cr 0&i&0\cr 0&0&i\cr}V_T
\pmatrix{1&0&0\cr 0&1 &0\cr 0&0&-i\cr}
=\frac{1}{\sqrt 3} 
\pmatrix{1&1&-i\cr i\z &i\z^2 &1\cr i\z^2 &i\z &1\cr}\;.
\ena
Therefore, the CP violation phase intrinsic to a
Dirac neutrino system is $\delta=\pi/2$, i.e., the 
maximal CP violation. There are two other phases that 
are intrinsic to Majorana neutrino system which is 
the same as the general case given in Eq.(15).
 
\vskip 5mm
 
\apsc{Other ansatz about mass parameters}

In the text, we considered the model which predicts 
$V_{2j}=V_{3j}^*$. Here we consider other such possibilities. 

\vskip 3mm
\noindent
(a) The mass matrix which predicts $V_{1j}=V_{3j}^*$

We consider the case that $m^0_1$ and $\tilde{m}_1$ 
are proportional to $\z^2$,
$m^0_2$ and $\tilde{m}_2$ to $\z$, and
$m^0_3$ and $\tilde{m}_3$ to 1. When this mass matrix is 
transformed by $V_T$, we obtain the mass matrix $\tilde m_\nu$ 
given in Eq.(10) which is a complex symmetric matrix. However, 
these complex phases are removed by the phase transformation 
by phase matrix ${\rm diag}(\z^2,\z,1)$ and 
$\tilde m_{\nu}$ can be transformed into real symmetric matrix. 
That is, by the tri-maximal mixing matrix 
\bea
V'_T = V_T\pmatrix{\z^2&0&0\cr 0&\z&0\cr 0&0&1\cr}
 \;,
\ena
the mass matrix $m_\nu$ is transformed by a real symmetric mass 
matrix $\tilde{m}_\nu'$
\bea
\tilde{m}_\nu' = V^{'T}_T m_{\nu} V'_T\;.
\ena
This mass matrix is diagonalized by an orthogonal matrix $O'$. 

Thus, the mixing matrix is given by 
\bea
V=V_T'O'\;.
\ena
This mixing matrix has the property 
that $V_{1j}=V_{3j}^*$ for $j = $ 1,2,3. As we discussed 
in the text, this condition implies that 
$|(V_{SF})_{1j}|=|(V_{SF})_{3j}|$ for $j = $ 1,2,3. 
By solving these equations, we find
\bea
c^2_{23} = \frac{s^2_{13}}{c^2_{13}}\;,\qquad
\cos \delta=-\frac{ s_{23} }{c_{23}s_{13}}\cot{ 2 \theta_{12} } \;.
\ena

Since the CHOOZ data gives the severe constraint, $s_{13}^2<0.05$ 
and $c_{23}^2\simeq s_{13}^2$, we can not predict the large 
mixing between $\nu_\mu$ and $\nu_\tau$. Thus, unfortunately this 
model can not explain the atmospheric data and the CHOOZ data 
simultaneously.

\vskip 3mm
\noindent
(b) The mass matrix which predicts $V_{1j}=V_{2j}^*$

We consider the case that  
$m^0_1$ and $\tilde{m}_1$ are proportional to $\z$,
$m^0_2$ and $\tilde{m}_2$ to $\z^2$, and
$m^0_3$ and $\tilde{m}_3$ to 1.
By repeating the same discussion for the previous case, 
we find that  $m_{\nu}$ can be transformed into real symmetric mass 
matrix $\tilde{m}_{\nu}$ by the tri-maximal mixing matrix as  
\bea
\tilde{m}_\nu = V''_T m_{\nu} V''_T\;,
\ena
where
\bea
V''_T =V_T \pmatrix{\z&0&0\cr 0&\z^2&0\cr 0&0&1\cr}\;.
\ena
Then, we find that $V_{1j}=V_{2j}^*$ for $j = $ 1,2,3, which 
implies that 
$|(V_{SF})_{1j}|=|(V_{SF})_{2j}|$ for $j = $ 1,2,3. 
We find
\bea
s^2_{23} = \frac{s^2_{13}}{c^2_{13}}\;,\qquad
\cos{\delta} = \frac{c_{23}}{s_{23} s_{13} }\cot{ 2 \theta_{12} }\;.
\ena

Since $s^2_{23}$ should be very small to explain the CHOOZ data, 
this model can not explain the atmospheric data.

\end{document}